\documentclass[
    ,final            
  ]
  {aipproc}

\layoutstyle{6x9}
\usepackage{epsf}

\begin{document}

\title{Wave Algorithms:\\ Optimal Database Search and Catalysis\footnote{%
Invited talk at the conference ``Quantum Computing: BackAction 2006",
IIT Kanpur, India, March 2006.}}

\classification{03.67.Lx, 34.10.+x, 46.40.-f}
\keywords      {Grover's database search algorithm, Catalysis, Resonance}

\author{Apoorva D. Patel}{
  address={Centre for High Energy Physics
           and Supercomputer Education and Research Centre\\
           Indian Institute of Science, Bangalore-560012, India}
}

\begin{abstract}
Grover's database search algorithm, although discovered in the context
of quantum computation, can be implemented using any physical system that
allows superposition of states. A physical realization of this algorithm is
described using coupled simple harmonic oscillators, which can be exactly
solved in both classical and quantum domains. Classical wave algorithms
are far more stable against decoherence compared to their quantum
counterparts. In addition to providing convenient demonstration models,
they may have a role in practical situations, such as catalysis.
\end{abstract}

\maketitle


\section{Computing with Waves}

Any physical system---with some initial state, some final state, and some
interaction in between---is a candidate for processing information. One only
needs to construct a suitable map between physical properties of the system
and abstract mathematical variables. Most of the development in computer
algorithms has been in the framework of ``particle-like" discrete digital
languages. It is known that ``wave-like" analogue computation can also be
carried out (e.g. using RLC circuits), but that has not been explored as
intensively. The obvious reason is that discrete variables allow a degree
of precision, by implementation of error correction procedures, that
continuous variables cannot provide. In addition, computational complexity
is believed to be the same for digital and analogue algorithms, so the
choice between the two is left to considerations of hardware stability.

With the advent of quantum computation, several quantum algorithms have
been discovered, which are superior to their counterparts based on Boolean
logic. Naturally, with both ``particle-like" and ``wave-like" behaviour
at their disposal, quantum algorithms cannot do any worse than their
classical counterparts. It is routinely stated that the simple parallelism
provided by superposition and interference of quantum states is the key
ingredient for the superiority of quantum algorithms. Now both superposition
and interference are generic features of wave dynamics, and it is worthwhile
to investigate the advantages they bring to an algorithm by exploring
implementations based on "wave-like" behaviour alone. Of course, the
classical wave implementations will be less efficient than the fully
quantum ones, but they are expected to be much more stable (in particular,
they have no entanglement and much weaker decoherence) and so may turn
out to be useful in specific situations. With this motivation, I study
in this work, Grover's database search algorithm~\cite{grover}---a
straightforward and yet versatile algorithm that allows a variety of
physical realisations.

\subsection{The Optimal Search Algorithm}

Database search is an elementary computational task, whose efficiency is
measured in terms of the number of queries one has to make to the database
in order to find the desired item. In the conventional formulation of the
problem, the query is a binary oracle (i.e. a Yes/No question). For an
unsorted database of $N$ items, using classical Boolean logic, one requires
on the average $\langle Q\rangle=N$ binary queries to locate the desired item.
The number of queries is reduced to $\langle Q\rangle=(N+1)/2$, if the search
process has a memory so that an item rejected once is not picked up again for
inspection.

Grover discovered a search algorithm that, using superposition of states,
reduces the number of required queries to $Q=O(\sqrt{N})$ \cite{grover}.
This algorithm starts with a superposition state, where each item has an
equal probability to get picked, and evolves it to a target state where
only the desired item can get picked. Following Dirac's notation, and using
the index $i$ to label the items, the starting and the target state satisfy
\begin{equation}
|\langle i|s \rangle|^2 = 1/N ~,~~ |\langle i|t \rangle|^2 = \delta_{it} ~~.
\end{equation}
The algorithm evolves $|s\rangle$ towards $|t\rangle$, by discrete rotations
in the two-dimensional space formed by $|s\rangle$ and $|t\rangle$, using
the two reflection operators,
\begin{equation}
U_t = 1 - 2|t\rangle\langle t| ~,~~ U_s = 1 - 2|s\rangle\langle s| ~~,
\end{equation}
\begin{equation}
(-U_sU_t)^Q |s\rangle = |t\rangle ~~.
\label{GroverAlgo}
\end{equation}
$U_t$ is the binary oracle which flips the sign of the target state
amplitude, while $-U_s$ performs the reflection-in-the-mean operation.
Solution to Eq.(\ref{GroverAlgo}) determines the number of queries as
\begin{equation}
(2Q+1) \sin^{-1} (1/\sqrt{N}) = \pi/2 ~~.
\label{querysoln}
\end{equation}
(In practice, $Q$ must be an integer, while Eq.(\ref{querysoln}) may not have
an integer solution. In such cases, the algorithm is stopped when the state
has evolved sufficiently close to, although not exactly equal to, $|t\rangle$.
Then one finds the desired item with a high probability.)

In the qubit implementation of the algorithm, one chooses $N=2^n$
and the items in the database are labeled with binary digits.
With the uniform superposition as the starting state,
\begin{equation}
\langle i|s \rangle = 1/\sqrt{N} ~,~~ 
U_s = H^{\otimes n}(1 - 2|0\rangle\langle 0|)H^{\otimes n} ~~,
\end{equation}
($H$ is the Hadamard operator), the implementation requires
only $O(\log_2 N)$ spatial resources \cite{grover}.
It has been proved that this is the optimal algorithm for
unsorted database search \cite{zalka}.

Different physical realizations of the database items (e.g. binary labels or
individual modes) and the target query oracle (e.g discrete binary operation
or continuous time evolution) produce a variety of implementations of this
algorithm. In the original version, the states are encoded in an $n$-qubit
register, and the oracle is a discrete binary operation (denoted by $U_t$
above). In the analogue version of the algorithm, the discrete unitary oracle
is traded for a continuous time interaction Hamiltonian, which evolves the
target state somewhat differently than the rest and acts for the entire
duration of the algorithm, and the number of queries is replaced by the
time one has to wait for before finding the target state \cite{farhi}.
The wave version of the algorithm requires $N$ distinct wave modes, instead of
$n$ qubits, but does not involve quantum entanglement at any stage \cite{lloyd}.
Such a wave search has been experimentally implemented using classical
Fourier optics, with a phase-shift plate providing the oracle \cite{spreeuw}.
An analogue version of the algorithm has been described using a classical
coupled pendulum model, where one of the pendulums is slightly different
than the rest and the uniform superposition state $|s\rangle$ is identified
with the center-of-mass mode \cite{anirvan}.
In what follows, I describe a binary oracle version of the wave search
algorithm using identical coupled harmonic oscillators.

\section{Harmonic Oscillator Implementation}

A harmonic oscillator is the favourite model of physicists. It provides
the first approximation in a wide variety of physical phenomena involving
small fluctuations about an equilibrium configuration. Its mathematical
description contains only quadratic forms, in position as well as momentum
coordinates. It can be solved exactly in both classical and quantum domains,
which makes it extremely useful in situations where a cross-over between
classical and quantum behaviour is to be analyzed. We shall first look at
the classical system, and then observe that the quantum system essentially
follows the same pattern.

\subsection{Classical Oscillators}

Let the items in the database be represented by $N$ identical harmonic
oscillators. While they are oscillating in a specific manner, someone taps
one of the oscillators (i.e. elastically reflects it by touching it).
The task is to identify which of the oscillators has been tapped, without
looking at the tapping---quite like a magician who can tell which one of
his cards was touched when he was blindfolded. The optimization criterion
is to design the system of oscillators, and their initial state, so as to
make the identification as quickly as possible.

\begin{figure}[b]
\epsfxsize=9truecm
\centerline{\epsfbox{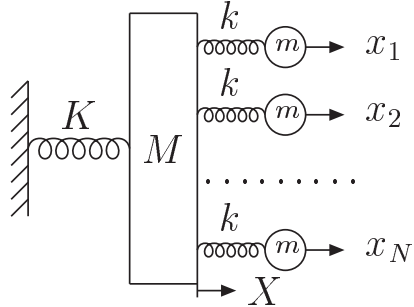}}
\caption{A system of $N$ identical harmonic oscillators,
coupled to a big oscillator via the center-of-mass mode.}
\end{figure}

Grover's algorithm requires identical coupling between any pair of
oscillators. This can be accomplished by coupling all the oscillators
to a big oscillator, as shown in Fig.1. The big oscillator is thus coupled
to the centre-of-mass mode, and becomes an intermediary between any pair of
oscillators with the same strength. The Lagrangian for the whole system is
\begin{equation}
{\cal L} = \frac{1}{2}M\dot{X}^2 - \frac{1}{2}KX^2
         + \sum_{i=1}^N [ \frac{1}{2}m\dot{x}_i^2 - \frac{1}{2}k(x_i-X)^2 ] ~.
\end{equation}
With the center-of-mass displacement, ${\overline x}\equiv\sum_{i=1}^N x_i/N$,
the Lagrangian can be rewritten as
\begin{equation}
{\cal L} = \frac{1}{2}M\dot{X}^2 - \frac{1}{2}KX^2
         + \frac{1}{2}Nm\dot{\overline x}^2 - \frac{1}{2}Nk({\overline x}-X)^2
         + \sum_{i=1}^N [ \frac{1}{2}m(\dot{x}_i-\dot{\overline x})^2
         -  \frac{1}{2}k(x_i-{\overline x})^2 ] ~.
\end{equation}

\begin{figure}[b]
\epsfxsize=12truecm
\centerline{\epsfbox{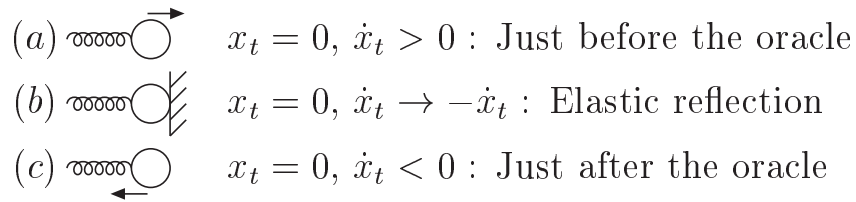}}
\caption{The binary tapping oracle flips the sign of the target oscillator
velocity, when its displacement is zero.}
\end{figure}

Now we can fix the oscillator parameters to implement Grover's algorithm.
In the algorithm, we are interested in the dynamics of the tapped oscillator.
All the other oscillators (i.e. $i \ne t$) influence the dynamics of $x_t$
only through the combination $\overline x$. The dynamics of $(N-2)$ linearly
independent modes orthogonal to $x_t$ and $\overline x$ (they all have the
form $(x_{j \ne t}-x_{k \ne t})$) decouples from the modes of interest;
we can drop them and effectively work in the 3-dimensional space of the
modes $\{X,{\overline x},x_t\}$. (In what follows, I shall first specify
initial conditions such that all $x_{i \ne t}$ are identical and all the
decoupled modes vanish. Subsequently, we will look at the general situation
by adding back all the decoupled modes.)

Choosing units of mass and time such that $m=1,k=1$, and in terms of the
variables
\begin{equation}
Y = \sqrt{M}X ~,~~
{\overline y} = \sqrt{N}{\overline x} ~,~~
y_t = x_t - {\overline x} ~,
\end{equation}
the effective Lagrangian becomes
\begin{equation}
{\cal L}_{\rm eff} = \frac{1}{2}\dot{Y}^2 - \frac{1}{2}\frac{K}{M}Y^2
                   + \frac{1}{2}\dot{\overline y}^2
                   - \frac{1}{2}\left({\overline y}-\sqrt{\frac{N}{M}}Y\right)^2
                   + \frac{N}{2(N-1)}\dot{y}_t^2 - \frac{N}{2(N-1)}y_t^2 ~.
\end{equation}
The potential energy terms in ${\cal L}_{\rm eff}$ are easily diagonalized,
and yield the eigenvalues
\begin{eqnarray}
&&\omega_\pm^2 = \frac{1}{2}\left(1+\frac{K+N}{M}\right)
       \pm \sqrt{\frac{1}{4}\left(1+\frac{K+N}{M}\right)^2 - \frac{K}{M}} ~,~~
\nonumber \\
&&\omega_+^2 + \omega_-^2 = 1 + \frac{K+N}{M} ~,~~
  \omega_+^2 \omega_-^2 = \frac{K}{M} ~,~~
  \omega_t = 1 ~~.
\end{eqnarray}
The corresponding eigenmodes are
\begin{eqnarray}
e_\pm &=& (1-\omega_\pm^2)Y + \sqrt{\frac{N}{M}}{\overline y}
       =  (1-\omega_\pm^2)\sqrt{M}X + \frac{N}{\sqrt{M}}{\overline x} ~,\\
e_t   &=& y_t = x_t - {\overline x} ~~.\nonumber
\end{eqnarray}

The initial uniform superposition state can be realized as all the
oscillators moving together, while the big oscillator is at rest.
\begin{equation}
t=0:\quad X=0 ~,~~ \dot{X}=0 ~,~~ x_i = 0 ~,~~ \dot{x}_i = A ~.
\end{equation}
(We will consider situations with general initial conditions later.)
The reflection operators correspond to shifting the appropriate oscillator
phases by half a period. The binary tapping oracle can be realized as the
elastic reflection illustrated in Fig.2. That implements $U_t$ in the
velocity space, by reversing the target oscillator velocity at the instance
when all the displacements vanish. Time evolution of the coupled oscillators
redistributes the total kinetic energy, and that can implement the operator
$U_s$ with a suitable choice of time interval and frequencies.

With the natural frequency of individual oscillators $\omega=\sqrt{k/m}=1$,
the reflection-in-the-mean operation requires $\omega_\pm$ to be rational
numbers. Optimization means that they should be selected to make the dynamics
of the whole system of oscillators have as small a period as possible.
The solution is not unique. One set of solutions is ($p$ is a positive integer)
\begin{equation}
\omega_+ = \frac{2p+1}{2} ~,~~ \omega_- = \frac{1}{2}
~~\Longrightarrow~~ M = \frac{16Nm}{3(2p+3)(2p-1)} ~,~~
                    K = \frac{(2p+1)^2 Nk}{3(2p+3)(2p-1)} ~.
\label{freqone}
\end{equation}
In these cases, in absence of oracles, the dynamics of the whole system
of oscillators has the period, $T=4\pi$. The big oscillator returns to
its initial rest state ($X=0, \dot{X}=0$), whenever $t$ is an integral
multiple of $2\pi$, i.e. after every half a period. Time evolution for
the same interval of half a period reverses $\dot{\overline x}$, while
leaving $\dot{x}_t - \dot{\overline x}$ unchanged, i.e. it implements
the operator $U_s$ in the velocity space. Thus Grover's algorithm,
Eq.(\ref{GroverAlgo}), can be realized by applying the tapping oracle
at every time interval $\Delta t=2\pi$.

A more interesting set of solutions is
\begin{equation}
\omega_+ = 2p ~,~~ \omega_- = 0
~~\Longrightarrow~~ M = \frac{Nm}{(2p+1)(2p-1)} ~,~~ K = 0 ~.
\label{freqtwo}
\end{equation}
Under these conditions, the big oscillator is not coupled to any support,
and $e_-$ becomes a translation mode for the whole system of oscillators.
The translation mode can be eliminated from the dynamics with the initial
conditions
\begin{equation}
t=0:\quad X=0 ~,~~ \dot{X}= -\frac{N}{M}A ~,~~ x_i = 0 ~,~~ \dot{x}_i = A ~.
\end{equation}
Then, in absence of oracles, the dynamics of the whole system of oscillators
has the smallest possible period, $T=2\pi$. After half a period, the big
oscillator is back to its initial state, $\dot{\overline x}$ also returns
to its initial value, while $\dot{x}_t - \dot{\overline x}$ changes its
sign. This is equivalent to applying $-U_s$ in the velocity space, and
Grover's algorithm can be implemented by tapping the target oscillator
at every time interval $\Delta t=\pi$.

There is an important physical distinction between the quantum and the
wave implementations of the amplitude amplification process in Grover's
algorithm---quantum probability is mapped to wave energy. The enhancement
of the quantum amplitude increases the probability of finding the target
state $N$-fold, while the enhancement of the wave amplitude increases the
energy of the target oscillator $N$-fold. The well-known phenomenon of
``beats'' is responsible for energy transfer amongst coupled oscillators.
The elastic reflection oracle does not change energy, and it is interesting
to observe that the oscillator which is obstructed by tapping picks up energy. 

\subsection{Stability Considerations}

Now we can look at the behaviour of the wave implementation under more
general circumstances. First consider the initial conditions. Despite
appearances, precise synchronization of oscillators is not an issue in
the algorithm, because of the explicit coupling to the center-of-mass mode.
For instance, the algorithm can be started off with an initial push to the
big oscillator, $\dot{X}=B, \dot{x}_i=0$, and the system of oscillators
would evolve to the stage $\dot{X}=0, \dot{x}_i=A$. Furthermore, any
arbitrary distribution of initial velocities of oscillators can be
accommodated in the analysis by bringing back the $(N-2)$ decoupled modes.
The decoupled modes have no effect whatsoever on the dynamics of the
$\{X,{\overline x},x_t\}$ modes. Consequently, they modify the algorithm
only to the extent that the energy amplification of the target oscillator
is limited to the initial energy present in the $\{X,{\overline x},x_t\}$
modes, instead of being $N$-fold. Explicitly, the maximum gain is
\begin{equation}
\left[ \left( N\dot{\overline x}^2 +
              \frac{N}{(N-1)}(\dot{x}_t - \dot{\overline x})^2 \right)
       \bigg/ \dot{x}_t^2 \right]_{t=0} ~,
\label{maxgain}
\end{equation}
which can be substantial for the generic situation where the initial
$\dot{x}_t$ and $\dot{\overline x}$ are of the same order of magnitude.

To extract the maximum gain, the algorithm must be stopped at a precise
instant (i.e after a precise number of tapping oracles $Q$); otherwise
the evolution continues in repetitive cycles. The state evolution in
Grover's algorithm is a rotation at a uniform rate in the two dimensional
$|s\rangle$-$|t\rangle$ subspace. The average overlap of the target state,
with the state $|q(Q)\rangle$ after $Q$ queries, is therefore
\begin{equation}
|\langle q(Q)|t \rangle|^2_{\rm av}
= \langle \sin^2 \theta \rangle_{\rm av} = 1/2 ~.
\end{equation}
Thus if the algorithm is stopped at a random instant, the energy gain
on the average is half of its maximum value in Eq.(\ref{maxgain})---which
can still be substantially larger than $1$.

Next consider the effect of damping. Damping for a harmonic oscillator has
to be analyzed using its equation of motion; it cannot be described using
a time-independent Lagrangian. The standard description is:
\begin{equation}
\ddot{x} + 2\gamma\dot{x} + \omega_0^2 x = 0 ~~\Longrightarrow~~
x = A e^{-\gamma t} \cos\left(\sqrt{\omega_0^2 - \gamma^2}~t + \phi\right) ~.
\end{equation}
The crucial ingredient in the algorithm is the coherence amongst the phases
of the oscillators. That is governed by the frequencies of the oscillators,
and is independent of the amplitudes. For a weakly damped oscillator, its
amplitude changes linearly with the damping coefficient, while its frequency
changes quadratically. The time evolution of the above implementation,
therefore, remains essentially unaffected if the oscillators experience a
small damping. The dominant effect is a decrease in the energy amplification
due to decaying amplitudes.

Among other variations, simultaneous scaling of masses and spring constants
of the oscillators (i.e. $m_i = \alpha_i m$ and $k_i = \alpha_i k$) does
not alter the algorithm at all, since the scale factors can be absorbed by
redefining $x_i$. One can also contemplate interchange of the initial and
the final states, since the algorithm is fully reversible. The evolution
would then run backwards, and the physical interpretation would be the
dissipation of a large initial energy of the target oscillator in to a
uniform distribution over all the coupled oscillators.

\subsection{Quantum Domain}

The dynamics of harmonic oscillators is simple enough to permit exact quantum
analysis as well. It is convenient to interpolate between classical and
quantum domains using the coherent state formulation (see for example,
Ref.~\cite{cohentannoudji}).
Coherent states are superpositions of the energy eigenstates, parametrized
by a single complex variable $\alpha$,
\begin{equation}
|\alpha\rangle = e^{-|\alpha|^2/2} \sum_n {\alpha^n \over \sqrt{n!}}
               ~ |n\rangle ~.
\end{equation}
They describe Gaussian wavepackets with minimal spread (i.e. displaced
versions of the ground state eigenfunction),
\begin{equation}
\Delta x = \sqrt{\hbar \over 2m\omega} ~,~~
\Delta p = \sqrt{m\hbar\omega \over 2} ~.
\end{equation}
A coherent state with the initial condition $\alpha(t=0)=\alpha_0$ has
energy $\hbar\omega(|\alpha_0|^2+{1\over2})$, and the centre of its
wavepacket performs the same simple harmonic motion as a classical
particle would,
\begin{eqnarray}
\alpha_0 e^{-i\omega t} = {\langle x\rangle(t) \over 2\Delta x}
                        +i{\langle p\rangle(t) \over 2\Delta p} ~.
\end{eqnarray}

\begin{figure}[hbt]
\epsfxsize=12truecm
\centerline{\epsfbox{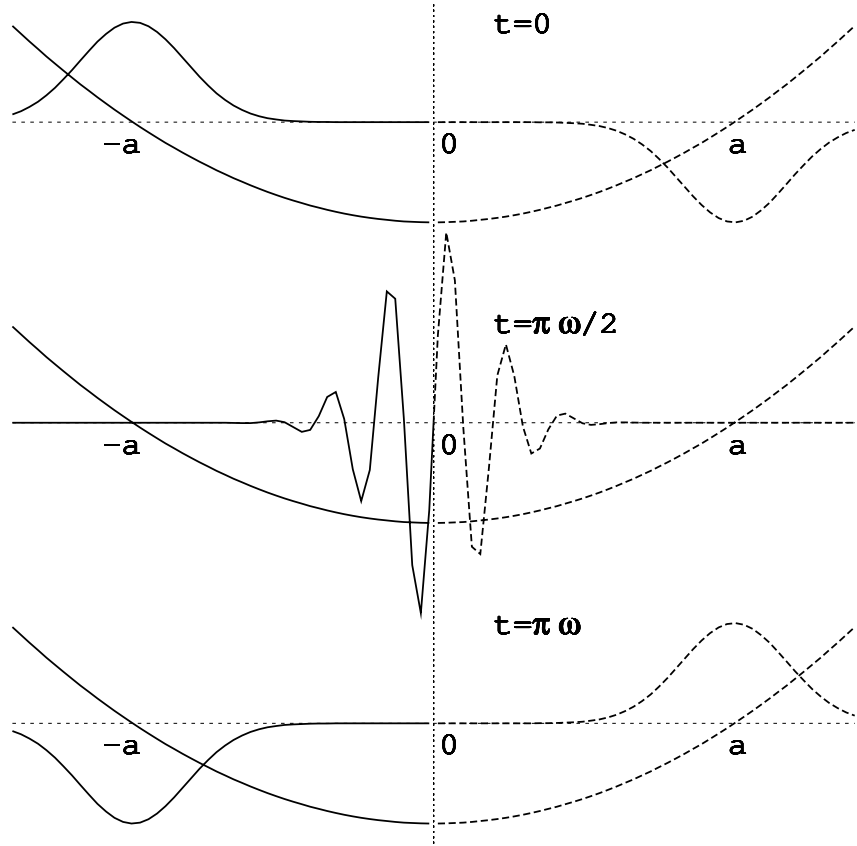}}
\caption{Evolution of the coherent state wavefunction of the tapped
oscillator, with the initial condition $\alpha_0=-a$, The left half of the
figure shows the actual wavepacket in the half-harmonic oscillator potential,
with the impenetrable wall at $x=0$. The right half of the figure shows
the image wavepacket that ensures the node of the wavefunction at $x=0$.
For $t=\pi\omega/2$, the wavefunction is purely imaginary, but the factor
of $i$ is omitted for convenience in drawing the figure. The wavepacket at
$t=\pi\omega$ includes the geometric phase of $-1$ arising from reflection.}
\end{figure}

The wavefunction of the state evolves according to
\begin{equation}
|\psi(0)\rangle = |\alpha_0\rangle ~~\Longrightarrow~~
|\psi(t)\rangle = e^{-i\omega t/2} |\alpha_0 e^{-i\omega t}\rangle ~,
\end{equation}
while the explicit structure of the wavepacket is given by
\begin{equation}
\psi(x) = \left( {m\omega \over \pi\hbar} \right)^{1/4}
          \exp\left[- \left( x-\langle x\rangle \over2 \Delta x \right)^2
                    + i{x\langle p\rangle \over \hbar} \right] ~.
\end{equation}

Thus the classical analysis of previous sections can be carried over
unchanged to the quantum domain, provided we can figure out how the tapping
oracle works for the coherent states. The tapped oscillator corresponds to
a particle moving in the half-harmonic oscillator potential
\begin{equation}
V(x) = \textstyle{1\over2} kx^2 ~~{\rm for}~ x\le0 ~,~~
V(x) = \infty ~~{\rm for}~ x>0 ~.
\end{equation}
The impenetrable wall at $x=0$ is equivalent to enforcing the boundary
condition $\psi(x=0)=0$. So the energy eigenstates of the half-harmonic
oscillator are the same as those for the harmonic oscillator, with odd $n$.
It is straightforward to ensure the node at $x=0$ using the method of images,
and the tapped oscillator coherent states become
\begin{equation}
|\alpha_t\rangle = C \big(|\alpha\rangle - |-\alpha\rangle\big) ~,
\end{equation}
with a time-independent normalization constant $C=(1-e^{-2|\alpha|^2})^{-1/2}$.
Tapping amounts to change-over between $|\alpha\rangle$ and $|-\alpha\rangle$,
which reverses $\langle x\rangle$ and $\langle p\rangle$ compared to the
untapped motion. In addition, the wavefunction changes sign, which is the
geometric phase corresponding to wave reflection. The evolution of a
coherent state wavepacket undergoing a reflection from the wall is depicted
in Fig.3.

\section{Possible Applications}

The oscillator based search process discussed above has the same algorithmic
efficiency as the proposals of Refs.\cite{anirvan,spreeuw,lloyd}---where it
differs from them is in the actual physical implementation. All these wave
implementations require exponentially more spatial resources compared to
their digital counterparts, $O(N)$ vs. $O(\log_2 N)$. On the other hand,
they reduce the number of oracle calls by exploiting superposition of states.
Note that no algorithm based on Boolean logic, either with serial or with
parallel implementation, can reduce the number of oracle calls for an
unsorted database search to less than $O(N)$.

Quantum algorithms are superior to wave algorithms, because they can use
superposition as well as reduce spatial resources. The reduction of spatial
resources, however, comes with the cost that quantum algorithms have to
work with entangled states. Quantum entanglement is far more sensitive to
decoherence caused by environmental disturbances than mere superposition,
and that has made physical implementations of quantum algorithms very
difficult. On the other hand, superposition of classical waves can be
fairly stable, even in presence of a small amount of damping, and that
can make wave implementations advantageous in specific physical contexts.

These features indicate that wave algorithms fall in a regime in between
classical and quantum algorithms---more efficient than the former and more
robust than the latter. They are likely to be useful in practical situations,
where $N$ is not very large and environmental disturbances are not negligible.
Indeed it is worthwhile to systematically explore them, just like randomized
algorithms have been \cite{motwani}.

In the specific case of the unsorted database search problem, the remarkable
simplicity of the oscillator implementation would allow construction of
convenient demonstration models, which can demonstrate the power of quantum
algorithms even at school level. Going further, the practically useful
property of the wave search algorithm is that it focuses energy of many
oscillator modes into one of them. So it is really interesting to think
of situations where the energy amplification process can have dramatic
consequences. One such possibility is very familiar to all of us---the
Boltzmann factor where the energy is in the exponent---and I describe below
a scenario for catalysis where involvement of new mechanisms can enhance
our understanding of the observed phenomena.

\subsection{Catalysis}

There exist a large number of chemical reactions which, although not
forbidden by energy conservation, are extremely slow because they have to
pass through an intermediate state of high energy. In these reactions,
the dominant term governing the reaction rate is the Boltzmann factor,
$\exp(-E_b/kT)$, with the barrier energy in the exponent. Only a tiny
fraction of the molecules in the tail of the thermal distribution are
energetic enough to go over the barrier and complete the reaction. It is
known that the rates of many such reactions can be enhanced by orders of
magnitude by adding suitable catalysts (enzymes in case of biochemical
reactions) to the reactants. The conventional explanation for the reaction
rate enhancement, called transition state theory, is that the catalysts
lower the energy of the intervening barrier by altering the chemical
environment of the reactants.

The preceding analysis of the wave search algorithm suggests another
mechanism for catalysis. Vibrations and rotations of molecules are
ubiquitous harmonic oscillator modes. The catalyst can act as the big
oscillator and transfer energy of many modes to the reactant which faces the
energy barrier. For example, the catalytic substrate can have many identical
molecules of one reactant stuck to it and vibrating, the second reactant
then comes drifting along and interacts with one of the stuck molecules,
that molecule picks up energy from its neighbours and the reaction gets
completed. The rate enhancement results not from lowering of the energy
barrier but from increase of the reactant energy. In such a scenario, for
maximum efficiency, the physical parameters (masses and spring constants)
need to have specific values. But even without perfectly tuned parameters,
there can be partial energy focusing that provides useful increase in the
reaction rate. Whether this mechanism exists among the known catalysts,
or whether we can design new type of catalysts that use this mechanism,
is open to investigation.

The contributions of the ``transition state" and the ``energy transfer"
mechanisms to a catalytic effect are not mutually exclusive. So it is
desirable to identify characteristics that can tell one of them apart
from another. I point out two observable features that can distinguish
the role of chemical environment from physical waves:\\
(1) Isotopic substitution in the reactants changes physical parameters
without altering chemical properties. The electronic potential is
essentially independent of the nuclear mass, and mass dependence enters
the conventional transition state theory only through diffusion and
tunneling effects. This mass dependence is rather weak and monotonic.
On the other hand, the nuclear mass strongly affects vibration and
rotation frequencies, which can substantially alter the energy transfer
amongst coupled oscillator modes. Also, resonant energy transfer is not
monotonic, i.e. it decreases on either side of the optimal parameter
value.\\
(2) In the transition state theory, the reaction takes place between
individual reactant molecules. On the other hand, energy transfer is a
cooperative phenomenon that cannot occur without participation of nearby
oscillator modes or molecules with similar properties. It would therefore
be enhanced by existence of non-reacting but similar chemical bonds close
to the reaction site, and also by increasing concentration of the reactants.

\subsection{Kinetic Isotope Effect}

In the context described above, isotope dependence of reaction rates is
a signal of involvement of physical (in contrast to chemical) features in
the catalytic process. Many examples of isotopic dependence of catalytic
reaction rates have been discovered, and the effect is referred to as the
``Westheimer effect'' or the ``kinetic isotope effect'' \cite{westheimer}.
The effect is the largest for substitution of hydrogen by deuterium or
tritium, and has been extensively studied for the rupture of C-H/C-D/C-T
bonds. The conventional transition state theory has been found inadequate
for a theoretical understanding in several cases involving enzymes, and
vibrationally enhanced quantum tunneling has been invoked as an alternative
\cite{VEGST}.

Specifically, the observed kinetic isotope effects are divided in to two
categories. The ``primary effect" results from isotopic substitution of
the hydrogen atom that is exchanged between the reactants. The ''secondary
effect" arises from isotopic substitution of a hydrogen atom that is not
exchanged during the reaction but is adjacent to the exchanged one.
Empirical models of reaction coordinate transition have been constructed,
but they require effects of quantum tunneling under the barrier as well
as contribution from vibrational bond energy to explain the size of the
observed effects. In particular, the secondary effect cannot be explained
at all without invoking coupled dynamics of atoms \cite{couplmotion}.
All these additional contributions enhance the reaction rates beyond
their values in the classical transition state theory, and hence would
be favoured in useful biochemical processes by natural selection during
evolution. The observed results, however, are quoted as ratios of reaction
rates to eliminate unknown normalization constants.

Against this backdrop, the coupled oscillator based energy transfer
mechanism described in earlier sections offers several novelties:\\
(i) The whole analysis is based on first principles. For $N$ coupled
oscillators, just a single reflection can amplify the vibrational energy
by a factor of $(3-\frac{4}{N})^2$, which can be as large as $9$.
Of course, the maximum amplification possible (with more reflections) is $N$.\\
(ii) The vibrational modes contributing to catalysis must be soft, i.e.
$\hbar\omega=O(kT)$, so that they can get thermally excited and participate
in the dynamics. This will produce a characteristic temperature dependence,
with the vibrational contribution dropping out at low enough temperatures.\\
(iii) Resonant energy transfer requires good frequency matching between 
coupled oscillator modes, as in Eqs.(\ref{freqone},\ref{freqtwo}). Some of
that can be inferred from the molecular structures (e.g. similar C-H bonds
involved in secondary kinetic isotope effect), and more can be tested by
spectroscopic methods.\\
(iv) The wavefunction sign-flip caused by reflection can switch between
bonding and anti-bonding molecular orbitals (see Fig.4), and thus help in
transfer of atoms. This feature is not related to either the mass or the
temperature.

Clearly, the coupled oscillator inspired catalytic mechanism needs to be
explored further, with careful modeling of specific reactions. It would
be an impressive achievement indeed, if the detailed understanding can be
used in design of new types of catalysts.

\begin{figure}[htb]
\epsfxsize=7.2truecm
\centerline{\epsfbox{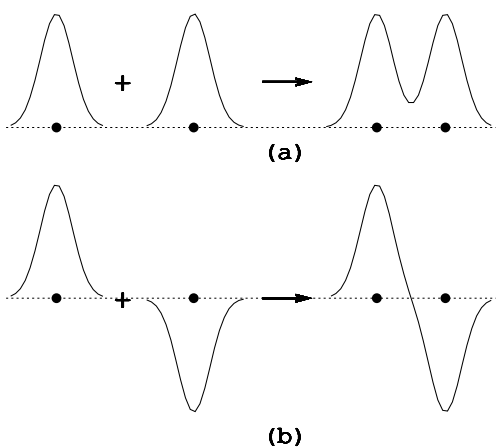}}
\caption{Formation of molecular bonds by overlap of electron clouds:
(a) a binding orbital is formed when the two wavefunctions are in phase,
(b) an anti-binding orbital results when the two wavefunctions have
opposite phases.} 
\end{figure}


\begin{theacknowledgments}
I thank the organizers for a wonderful conference that brought together
experimentalists and theorists working on a wide spread of topics related
to quantum computing.
Part of the work presented here has appeared elsewhere~\cite{wavsrch}.
\end{theacknowledgments}

\end{document}